\documentclass[useAMS,usenatbib]{mn2e}
\usepackage{psfig}
\usepackage[a4paper]{hyperref}
\usepackage{amssymb}
\usepackage{aas_macros}

\newcommand{\rmsub}[2]{#1_{\rm #2}} 

\title[The transiting planet WASP-3b]{WASP-3b: a strongly-irradiated transiting gas-giant planet}
\author[D. Pollacco et al.]
{
D. Pollacco$^{1}$\thanks{E-mail: d.pollacco@qub.ac.uk},
I. Skillen$^{7}$,
A. Collier Cameron$^{2}$,
B. Loeillet$^{11}$,
H.C. Stempels$^{2}$,
\newauthor
F. Bouchy$^{12,13}$,
N.P. Gibson$^{1}$,
L. Hebb$^{2}$,
G. H\'ebrard$^{12}$,
Y.C. Joshi$^{1}$,
I. McDonald$^{5}$,
\newauthor
B. Smalley$^{5}$,
A.M.S. Smith$^{2}$,
R.A. Street$^{1,15}$,
S. Udry$^{10}$,
R.G. West$^{3}$,
D.M. Wilson$^{5}$,
\newauthor
P.J. Wheatley$^{9}$
S. Aigrain$^{6}$,
C.R. Benn$^{7}$,
V.A. Bruce$^{2}$,
D.J. Christian$^{1}$,
W.I. Clarkson$^{4,14}$,
\newauthor
B. Enoch$^{4}$,
A. Evans$^{5}$,
A. Fitzsimmons$^{1}$,
C.A. Haswell$^{4}$,
C. Hellier$^{5}$,
S. Hickey$^{7,16}$,
\newauthor
S.T. Hodgkin$^{6}$,
K. Horne$^{2}$,
M. Hrudkov\'{a}$^{7,17}$,
J. Irwin$^{6}$,
S.R. Kane$^{7}$,
F.P. Keenan$^{1}$,
\newauthor
T.A. Lister$^{2,5,15}$,
P. Maxted$^{5}$,
M. Mayor$^{10}$,
C. Moutou$^{11}$,
A.J. Norton$^{4}$,
J. P. Osborne$^{3}$,
\newauthor
N. Parley$^{4}$,
F. Pont$^{10}$,
D. Queloz$^{10}$,
R. Ryans$^{1}$,
and
E. Simpson$^{1}$
\\
\\
$^{1}$Astrophysics Research Centre, School of Mathematics \&\ Physics, Queen's University, University Road, Belfast, BT7 1NN, UK\\
$^{2}$School of Physics and Astronomy, University of St Andrews, North Haugh, St Andrews, Fife KY16 9SS, UK\\
$^{3}$Department of Physics and Astronomy, University of Leicester, Leicester, LE1 7RH, UK\\
$^{4}$Department of Physics and Astronomy, The Open University, Milton Keynes, MK7 6AA, UK\\
$^{5}$Astrophysics Group, Keele University, Staffordshire, ST5 5BG\\
$^{6}$Institute of Astronomy, University of Cambridge, Madingley Road, Cambridge, CB3 0HA, UK\\
$^{7}$Isaac Newton Group of Telescopes, Apartado de Correos 321, E-38700 Santa Cruz de la Palma, Tenerife, Spain \\
$^{8}$Department of Astronomy, University of Florida, 211 Bryant Space Science Center, Gainesville, FL 32611-2055, USA\\
$^{9}$Department of Physics, University of Warwick, Coventry CV4 7AL, UK\\
$^{10}$Observatoire de Gen\`eve, Universit\'e de Gen\`eve, 51 Ch. des Maillettes, 1290 Sauverny, Switzerland\\
$^{11}$Laboratoire d'Astrophysique de Marseille, BP 8, 13376 Marseille Cedex 12, France\\
$^{12}$Institut d'Astrophysique de Paris, CNRS (UMR 7095) --  Universit\'e Pierre \&\ Marie Curie, 98$^{bis}$ bvd. Arago, 75014 Paris, France\\
$^{13}$Observatoire de Haute-Provence, 04870 St Michel l'Observatoire, France\\
$^{14}$STScI, 3700 San Martin Drive, Baltimore, MD 21218, USA\\
$^{15}$Las Cumbres Observatory, 6740 Cortona Dr. Suite 102, Santa Barbara, CA 93117, USA\\
$^{16}$Centre for Astrophysics, Science \& Technology Research Institute, University of Hertfordshire, Hatfield, AL10 9AB, UK \\
$^{17}$Astronomical Institute, Charles University Prague, V Holesovickach 2, CZ-180 00 Praha, Czech Republic\\
}

\begin{document}

\date{Accepted 1988 December 15. Received 1988 December 14; in original form 1988 October 11}

\pagerange{\pageref{firstpage}--\pageref{lastpage}} \pubyear{2002}

\maketitle

\label{firstpage}

\begin{abstract}

We report the discovery of WASP-3b, the third transiting 
exoplanet to be discovered by the WASP and SOPHIE collaboration. WASP-3b
transits its  host star USNO-B1.0 1256-0285133 every 
$ 1.846834 \pm 0.000002  $ days. Our
high precision radial-velocity measurements present a variation with
amplitude characteristic of a planetary-mass companion and 
in-phase with the light-curve. Adaptive optics imaging shows no
evidence for nearby stellar companions, and line-bisector analysis 
excludes faint, unresolved binarity and stellar activity as the cause 
of the radial-velocity variations. We make a preliminary 
spectroscopic analysis of the host star finding it to have $T_{\rm eff} = 6400 \pm 100$\,K 
and $\log g = 4.25 \pm 0.05$ which suggests it is most likely an
unevolved main sequence star of spectral type F7-8V. Our simultaneous 
modelling of the transit photometry and reflex motion of the host 
leads us to derive a mass of $1.76 ^{+ 0.08 }_{- 0.14 } M_J$  and radius 
$1.31 ^{+ 0.07 }_{- 0.14 } R_J$ for WASP-3b. The proximity and relative temperature of the host star
suggests that WASP-3b is one of the hottest exoplanets known, and thus has the potential to place stringent constraints on exoplanet atmospheric models.

\end{abstract}


\section{Introduction}

Since the discovery by \citet{henry2000} and \citet{c5} of the first exoplanet found to transit its host star, HD209458b, a further 22 transiting systems have been announced (see http://obswww.unige.ch/\~{}pont/TRANSITS.htm). Transiting exoplanets are highly prized because the transit geometry constrains the orbital inclination, and this in turn allows their masses and radii to be determined directly. The mass-radius relation for exoplanets allows us to probe their internal structure, since it is these parameters which are compared with models of planetary structure and evolution \citep{sato2005, guillot2006}. The limited numbers of transiting exoplanets studied so far show remarkable diversity in their physical parameters. For example, planets with masses $M\sim 1M_J$ range in size from 0.8--1.5$R_J$ for reasons that still elude us, although several plausible explanations  have been proposed \citep{arras2006, burrows2007, fortney2007, gu2004}. The discovery of transiting planets in greater numbers will allow us to further explore the mass-radius plane, and thereby constrain theories of planetary formation, migration and evolution.

The discovery of the first exoplanet, 51\,Peg\,b, was by the radial-velocity  method \citep{mq1}, and this technique is responsible for the discovery of the vast  majority of the known exoplanetary systems, including HD209458b. However, following the discovery of the transits of HD209458b \citep{henry2000, c5} it was  widely believed that the multiplex advantage of wide-field photometric imaging could lead to this technique becoming the dominant method for detecting exoplanets.  While it is true to say that, at least initially, photometric surveys have been slow to realise their expected detection rates \citep{h1}, recently
this situation has begun to change with 14 new systems published in 2006-07 alone. This improved detection rate is largely due to the development of a better understanding of noise characteristics, especially the correlated noise inherent in such photometric surveys  \citep{p2}. Transiting planets now comprise approximately 10\% of the known exoplanets.  Successful exoplanet photometric surveys include Wide Angle Search for Planets (WASP) Project  \citep{p1}, the Hungarian Automatic Telescope  (HAT) Network \citep{b2}, OGLE \citep{ogle}, the Transatlantic Exoplanet (TrES) Survey \citep{d1,o1}  and the XO group \citep{m1}. The WASP project has published two new systems in the last year WASP-1b and WASP-2b \citep{c4} and  WASP-1b, in particular, has proved to be especially interesting, being well oversized compared to other planets for its mass.

In this paper the WASP and SOPHIE collaboration announce the discovery of a new, relatively high mass, strongly-irradiated gas-giant exoplanet, WASP-3b. 

\section{Observations and data reduction}

\begin{figure}
\begin{center}
\psfig{figure=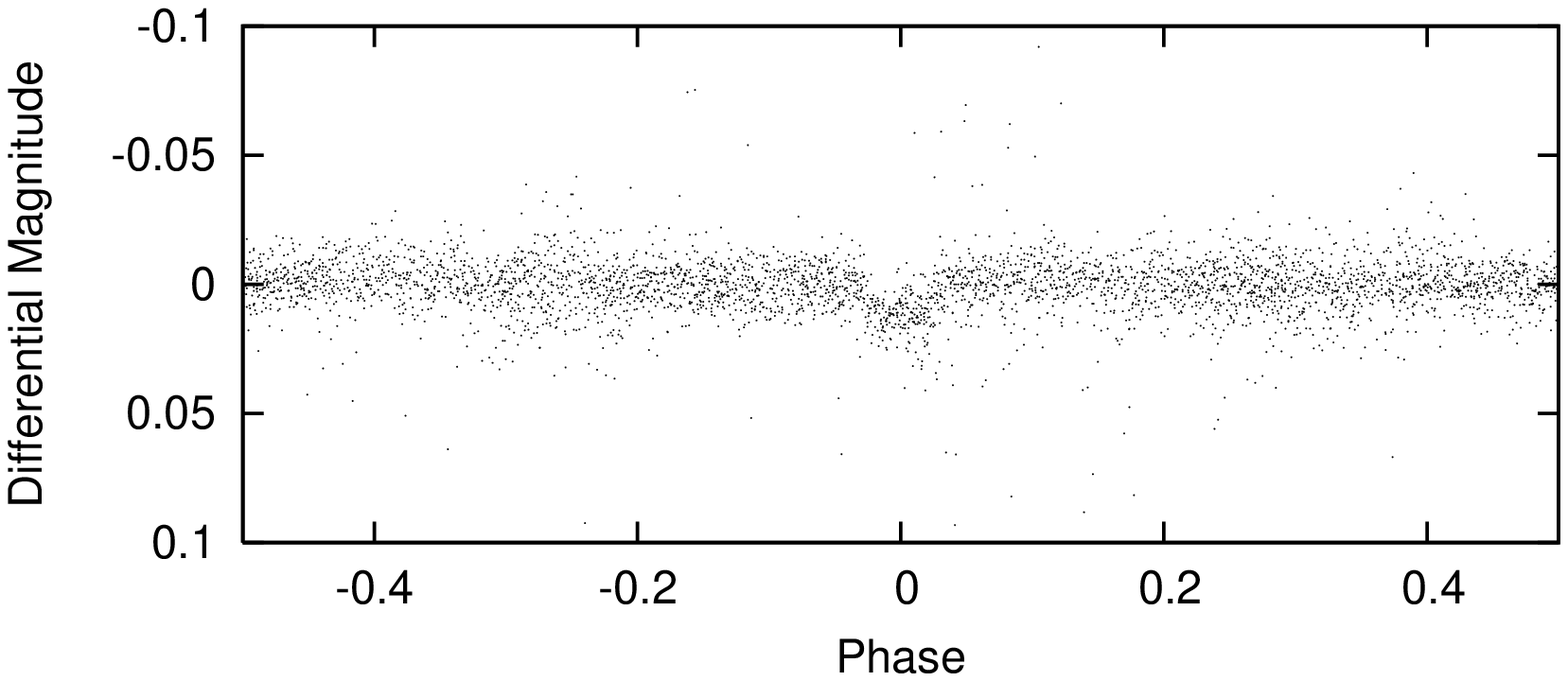,width=8.5cm}  

\psfig{figure=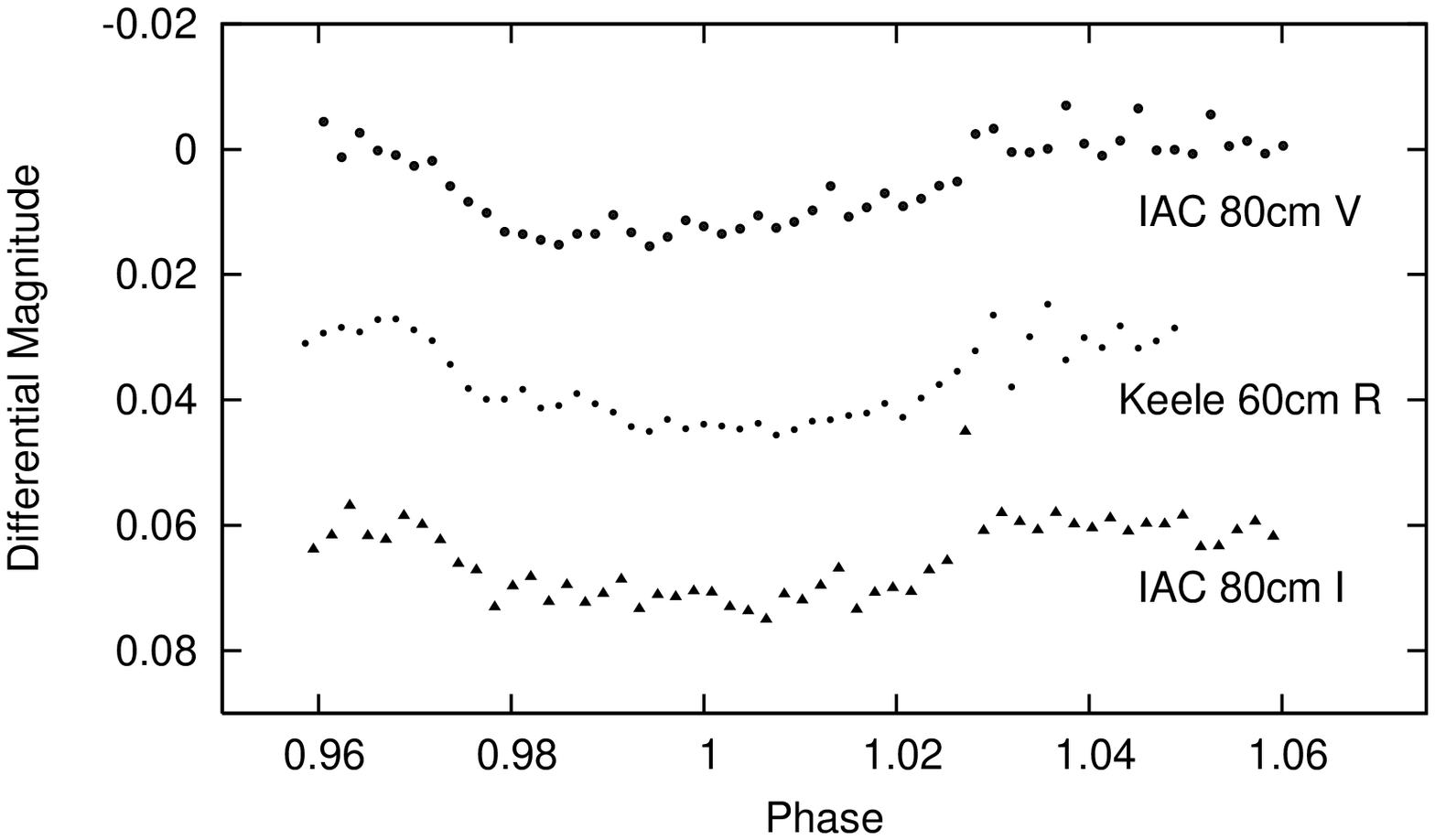,width=8.5cm} 
\caption[]{Light Curves for 1SWASP J183431.62+353941.4 (WASP-3) obtained with SuperWASP-N (top panel), the IAC\,80-cm telescope (V and I) and the Keele 60-cm telescope (R) as marked. All the data (apart from that from SuperWASP-N) was averaged in 300\,second bins. The data was phased using the ephemeris derived in Section~\ref{section:mcmc}, $T_0 = 2454143.8504$ and $P = 1.846834$ days.
}
\label{fig:lc}
\end{center}
\end{figure}

\subsection{SuperWASP-N Photometry}

The photometric observations used in this study are from the inaugural 2004 SuperWASP-N observing season, which ran from April to September of that year; this data set also led to the discovery of  WASP-1b and WASP-2b. Briefly, in 2004 the SuperWASP-N instrument comprised 4, and at times, 5 optical cameras, each consisting of a Canon 200\,mm f/1.8 telephoto lens imaging onto a  thermoelectrically-cooled, science-grade $2048 \times 2048$ CCD  camera (manufactured by e2v Technologies PLC). In this system, the CCD's 13.5 $\mu$m pixels project to an angular size of 14.2 arcseconds. For the entirety of the 2004 season, while robotic operation was being commissioned, the instrument performance was supervised with an observer always in attendance, although data acquisition was fully automated. Data were shipped to the UK on a weekly basis and reduced at the home institutes of the WASP Consortium using a dedicated, purpose-built pipeline, and the results ingested into the project database at the University of Leicester. The entire project infrastructure is described in detail by \citet{p1} along with the deployment of a further facility, WASP-S, at the South African Astronomical Observatory.  SuperWASP-N now runs completely robotically, and data are transferred to the UK in near real-time over the Internet. 
 
Transit searches were carried out on this dataset \citep{c1,c2,l1,s1} using the techniques outlined by \citet{c3}. 1SWASP J183431.62+353941.4, which we henceforth denote as WASP-3, was highlighted by \citet{s1} as a high-priority candidate worthy of further study. Figure~\ref{fig:lc} (top panel) shows the original SuperWASP-N lightcurve, which comprises 3969 data points obtained over a 118 day period.  In the original SuperWASP-N photometry 17 transits were observed with $>$50\% of a transit  observed on 10 ocassions. These data led to an ephemeris of $T_o$=2453139.1748 and $P$=1.846800 which was used to arrange followup observations. The transit here has a depth of 0.013 mag. and is 137 minutes in duration.

\subsection{Higher precision photometric observations}

WASP-3 was observed with the IAC\,80\,cm telescope as part of the Canarian Observatories' {\it International Time Programme} for 2007. The imaging camera on this telescope has an e2v Technology PLC CCD of $2148 \times 2148$  pixels giving a scale of 0.33 arcseconds/pixel and a total field of view of 10.6 arcminutes. Observations were taken during the transit of 2007 August 4, and consist of 327 images of 30 and 20 seconds integration in the $V$  and $I$ bands respectively. This night was photometric but suffered from significant Saharan dust extinction, estimated  to be $\sim 0.4$\,mag  on La Palma from the SuperWASP-N real-time pipeline. 

The images were bias subtracted with a stacked bias frame and flat-fielded with a stacked twilight  flat field image obtained in both filters using individual flats gathered over the course of the run.  After the instrumental signatures were removed, source detection and aperture photometry were performed on all science frames using the CASU catalogue extraction software \citep{il2001}.  We chose an aperture size 
matched to the typical seeing (5 pixels, 1.5$^{\prime\prime}$) and selected 5 non-variable comparison stars in the field of WASP-3 to use in deriving  the differential photometry.   For each exposure, we summed the fluxes of the 5 comparison stars  and divided by the flux of the target star to derive the differential magnitude of the target.  The resulting $V$ and $I$ band lightcurves (Figure~\ref{fig:lc}) of WASP-3 have a precision of $\sim 4$~millimag.

Further observations of WASP-3 were made with the Keele University Observatory 60cm Thornton Reflector on 2007 September 10. This telescope is equipped with a  765 $\times$ 510 pixel Santa Barbara Instrument Group (SBIG) ST7 CCD at the f/4.5 Newtonian focus, giving a 0.68 arcsecond/pixel resolution and a 8.63$\times$ 5.75 arcminute field of view. During most of the period the weather was photometric except post egress where some cloud appeared. Altogether 644$\times$20 sec observations in the $R$ band were obtained. After applying corrections for bias, dark current   and flat fielding in the usual way, aperture photometry on two comparisons were performed using the commercial software AIP4Win \citep{berry2005}.

Tracking errors and spurious electronic noise mean that systematic noise is  introduced into the system at an estimated level of  2~millimag  with periodicities of 2 and $\sim$20 minutes. No corrections have been applied for this effect.

\subsection{OHP 1.9\,m and SOPHIE}
\label{section:sophie}
\begin{table*}
\caption[]{Journal of radial-velocity measurements of WASP-3. The 1SWASP identifiers give the J2000 stellar coordinates of the photometric apertures; the USNO-B1.0 number denotes the star for which the radial-velocity measurements were secured.  The quoted uncertainties in the radial velocity errors include components due to photon noise (Section~\ref{section:sophie}) and 10m/s of jitter (Section~\ref{section:mcmc}) added in quadrature. The fourth and fifth columns give the FWHM of the CCF dip and the contrast of the dip as a fraction of the weighted mean continuum level. The signal-to-noise ratio at 550nm is given in column six.}
\label{tab:logspec}
\begin{center}
\begin{tabular}{cccccclcl}
BJD & $\rmsub{t}{exp}$ &$\rmsub{V}{r}$ & FWHM & Contrast & S:N & Notes \\
       &           (s)          &  km s$^{-1}$    & km s$^{-1}$ & \% &         & \\
\hline\\
\multicolumn{8}{l}{\bf 1SWASP J183431.62+353941.4: USNO-B1.0 1256-0285133 = GSC 02636-00195  = WASP-3 } \\ \\
2454286.5225	&  900   & $-5.751 \pm 0.018$   & 20.5 &  11.7	& 44 & Photometric \\
2454287.4563	&  900   & $-5.254 \pm 0.020$   & 19.8 &  11.8	& 37 & Cloud and Moonlight\\
2454289.3662	&  900   & $-5.259 \pm 0.018$   & 19.9 &  11.8	& 43 & Photometric and Moonlight \\
2454340.3251	& 1800  & $-5.648 \pm 0.013$   & 19.7 &  12.2	& 75 & Moonlight \\
2454341.3989	&  900   & $-5.406 \pm 0.015$   & 19.7 &  12.1	& 53 & Moonlight \\
2454342.3198 &  900   & $-5.544 \pm 0.019$  & 19.8  &  12.0   &  39  & Cloud and Moonlight \\
2454343.4825 &  2100  &  $-5.638 \pm 0.013$ & 19.8 & 12.2   & 75  & Cloud and Moonlight\\
\hline\\
\end{tabular}
\end{center}
\label{default}
\end{table*}

WASP-3 was observed with the Observatoire de Haute-Provence's 1.93\,m telescope and the SOPHIE 
spectrograph \citep{b1}, over the 8 nights 2007 July 2 -- 5 and August 27 -- 30; a total of 
7 usable spectra were acquired. SOPHIE is an environmentally stabilized spectrograph 
designed to give long-term stability at the level of a few m\,s$^{-1}$. We used the instrument in its 
high efficiency mode, acquiring simultaneous star and sky spectra through separate fibres
with a resolution of R=40000. Thorium-Argon calibration images were taken at the start and end 
of each night, and at 2- to 3-hourly intervals throughout the night.  The radial-velocity drift never exceeded 2-3 m/s, even on a night-to-night basis. 


Conditions during both runs varied from photometric to cloudy, but all nights were affected by 
strong moonlight. As WASP-3 has magnitude $V\sim10.5$, integrations of 900\,sec give a peak 
signal-to-noise per resolution element of around 40-50. The 2MASS colours and reduced proper 
motion for WASP-3 suggest a spectral type of about F7-8V, hence we cross-correlated the spectra 
against a G2V template provided by the SOPHIE control and reduction software. 

In all spectra the cross-correlation functions (CCF) were contaminated by the strong moonlight. We corrected them by using the CCF from the background light's spectrum (mostly the Moon) in the sky fibre. We then scaled both CCFs using the difference of efficiency between the two fibres.  Finally we subtracted the corresponding CCF of the background light from the star fibre, and fitted the resulting function by a Gaussian. The parameters obtained allow us to compute the photon-noise uncertainty of the corrected radial velocity measurement ($\sigma_{RV}$), using the relation detailed in \citet{c4}: 
$$
\sigma_{RV} = 1.7 * \sqrt{(FWHM)}/(SNR*Contrast)
$$
Overall our RV measurements have an average photon-noise uncertainty of 14 m/s. As our radial velocity measurements are not photon-noise limited, we quadratically added a radial velocity component to those uncertainties of about 10 m/s (more details in Section 3.2.1). The log of the observations and barycentric RV is given in Table 1.

\section{Results and Analyses}

\subsection{Stellar parameters}

\begin{figure*}
\begin{center}
\psfig{figure=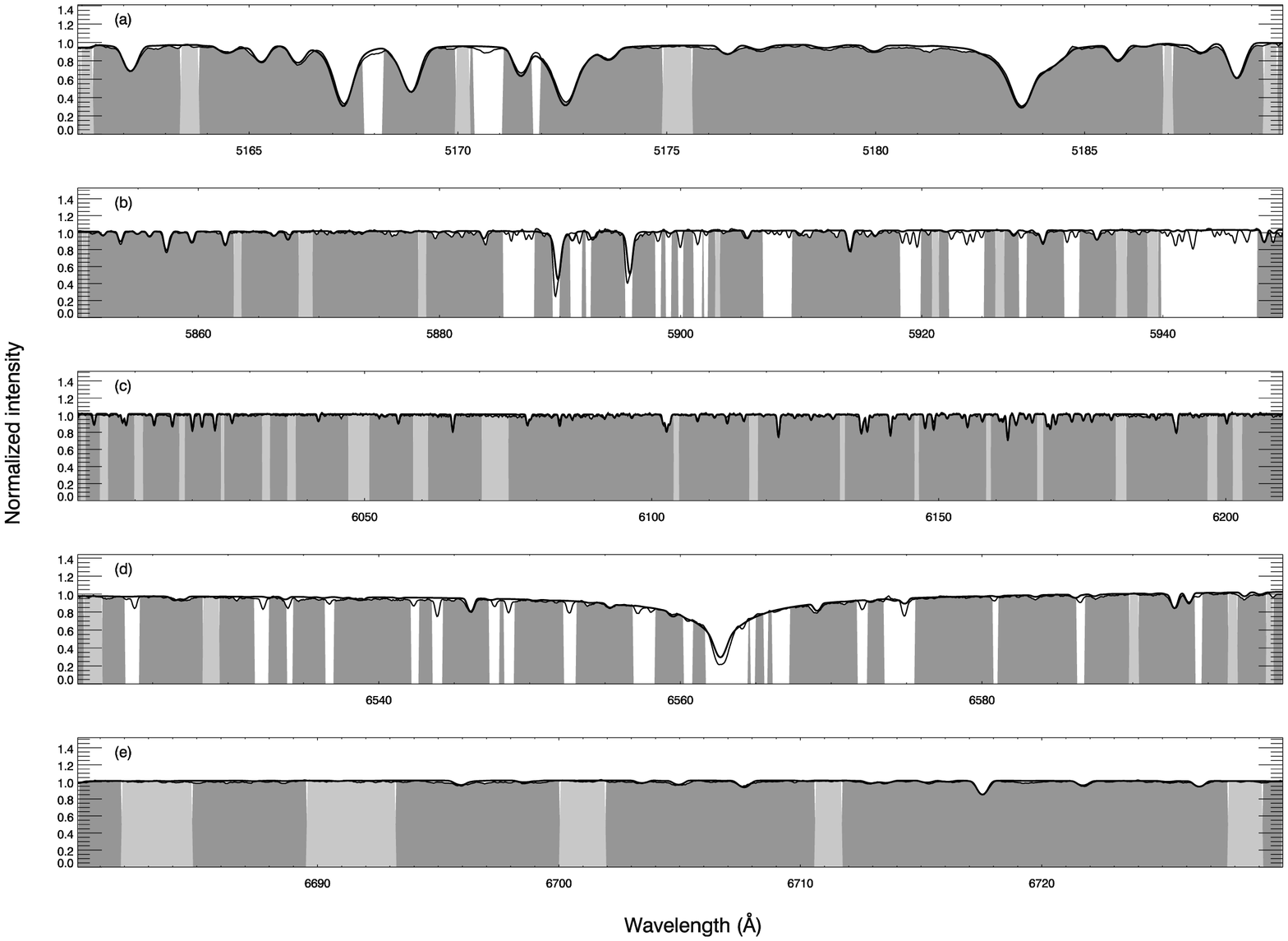,width=18cm,angle=90}
\caption[]{
A comparison between the observed SOPHIE spectrum of WASP-3 and the  calculated spectrum obtained from spectral synthesis with SME, using the atmospheric parameters from Table~\ref{tab:stellar_params}. The white regions are excluded from the spectral analysis, mainly because of the presence of telluric absorption. Light shaded regions were used to determine the continuum level, and the remaining dark shaded regions to determine the stellar atmospheric parameters. All five spectral sections were used simultaneously. The five sections contain (a) the Mg {\sc i} b triplet at 5175\AA\/ (b) the Na {\sc i} D doublet at 5890\AA/\ (c) a large region with well-isolated lines of a wide range of metals (d) the H$\alpha$ line at 6563\AA/\ and (e) the Li {\i} line at 6707\AA.
}
\label{fig:spec_synth}
\end{center}
\end{figure*}

The SOPHIE spectra are individually of modest signal-to-noise, but when summed together they are suitable for a preliminary photospheric analysis of WASP-3. However, from experience we have found that the SOPHIE standard pipeline reduction does not fully remove the scattered light component within
the spectrograph. While this does not affect radial velocities significantly, it can nonetheless have subtle effects on absorption line depths, adversely affecting the derived spectral synthesis parameters. Therefore we carefully re-reduced the first three raw images taken over 2--5 July 2007 with the {\sc reduce} echelle data reduction package \citep{pv2002},  paying careful attention to the issue of scattered light. These data are least affected by moonlight.

Following our analysis of WASP-1 \citep{s2}, we employed the methodology of \citet{v2}, using the same tools, techniques and model atmosphere grid. We used the {\sc idl}-based software {\it Spectroscopy Made Easy} ({\sc sme})  \citep{v1} to calculate and fit synthetic spectra using a multi-dimensional least squares approach.

We concentrated our analysis on five regions in the spectrum  (see Figure~\ref{fig:spec_synth}). These regions allow us to constrain the stellar effective temperature, $T_{\rm eff}$, (through the broad wings of H$\alpha$ and, to a lesser extent, Na {\sc i} D 5890\AA), gravity, $\log g$, (through Mg {\sc i} b 5175\AA\/ and Na {\sc i} D 5890\AA) and the metallicity, [M/H], (through the weak photospheric absorption lines in the 6000--6200\AA\/ region). We also measured the abundance of lithium from the Li {\i} 6708\AA\/ line. The combined spectrum is not of sufficient quality to perform a detailed abundance analysis.  The parameters we obtained from this analysis are listed in Table~\ref{tab:stellar_params}, and a comparison between observed and synthesized profiles is shown in Figure~\ref{fig:spec_synth}. In addition to the spectrum analysis, we also used Tycho $B$ and $V$, and 2MASS photometry to estimate the effective temperature using the Infrared Flux Method \citep{bs1}.  This gives $T_{\rm eff}  = 6200 \pm 300$ K, which is in close agreement with that obtained from the spectroscopic analysis. The Tycho and 2MASS colours ($V - K = 1.32$, $V - H = 0.2$) also suggest a spectral type of F7-8V  \citep{s1, c3}.

\begin{table}
\caption[]{Parameters for WASP-3 as derived from the SME analysis of the SOPHIE spectroscopy.}
\label{tab:stellar_params}
\begin{center}
\begin{tabular}{cc}
 Parameter    & WASP-3 \\
 \hline\\
$T_{\rm eff}$         & $6400 \pm 100$ K \\
log\,$g$  & $4.25   \pm 0.05$     \\
$[$M/H$]$  & $0.00  \pm 0.20$  \\ 
$v$\,sin\,$i$   & $13.4 \pm 1.5$ km/s          \\
$v_{\rm rad}$    &  -5.490 km/s         \\
\hline\\
\end{tabular}
\end{center}
\end{table}

In our spectra the Li {\sc i} $\lambda$6708\AA\/ line is weak, but still measurable, and we derive a Lithium abundance of $\log n({\rm Li}) + 12  = 2.0-2.5$. However, at this stellar temperature it is thought
that stellar age does not correlate well with Lithium abundance \citep{s3}, so  we have examined the evolutionary tracks for low and intermediate-mass stars presented by \citet{girardi2000} using a maximum-likelihood fitting routine. Using  our derived  stellar parameters we find the stellar mass $M_* = 1.24 \pm 0.08$ with an age of $0.7 - 3.5$\,Gyr.



\subsection{The reflex motion of the host star}

\subsubsection{Markov-chain Monte Carlo analysis}
\label{section:mcmc}

\begin{figure}
\begin{center}
\psfig{figure=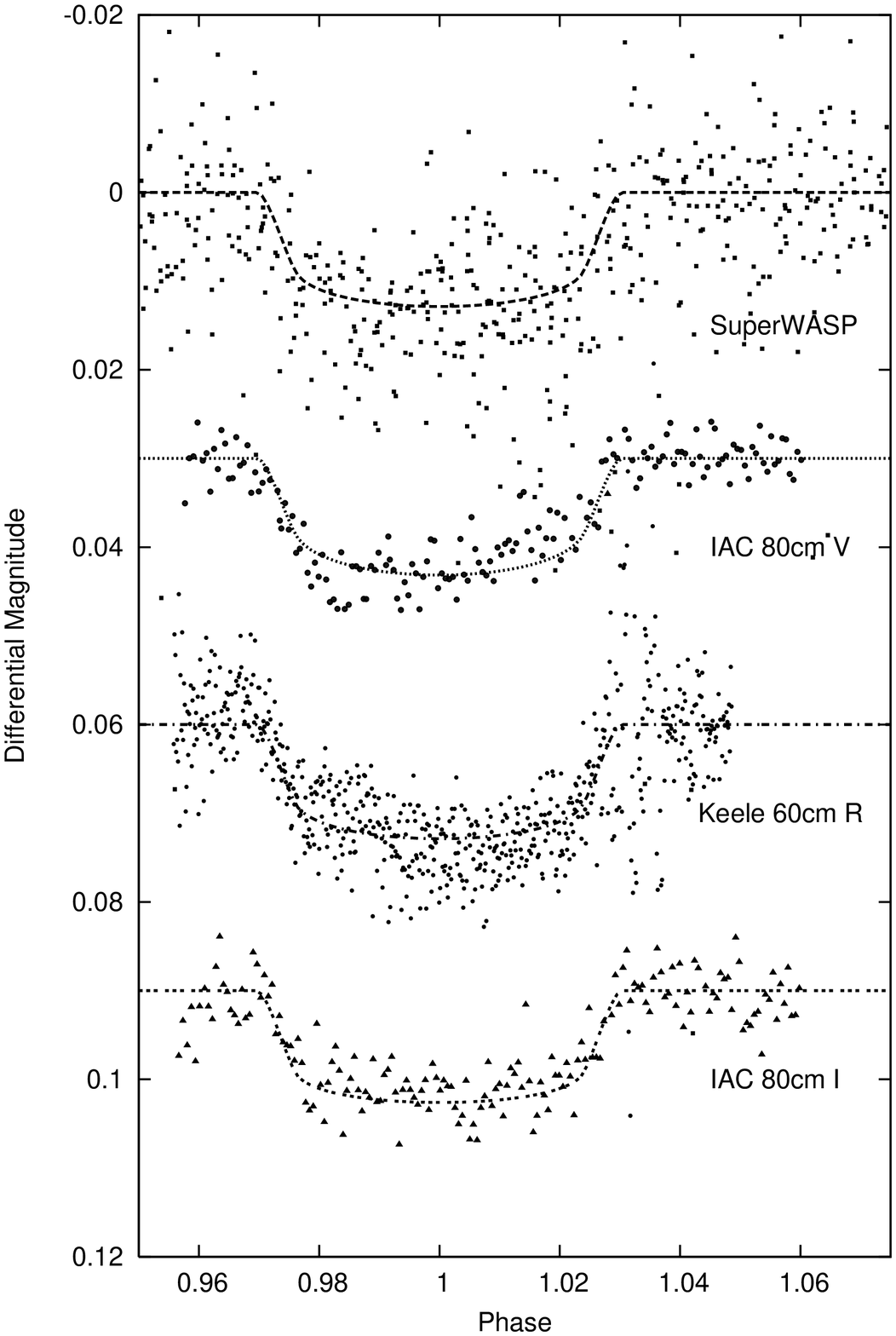,width=9.0cm,angle=0}

\psfig{figure=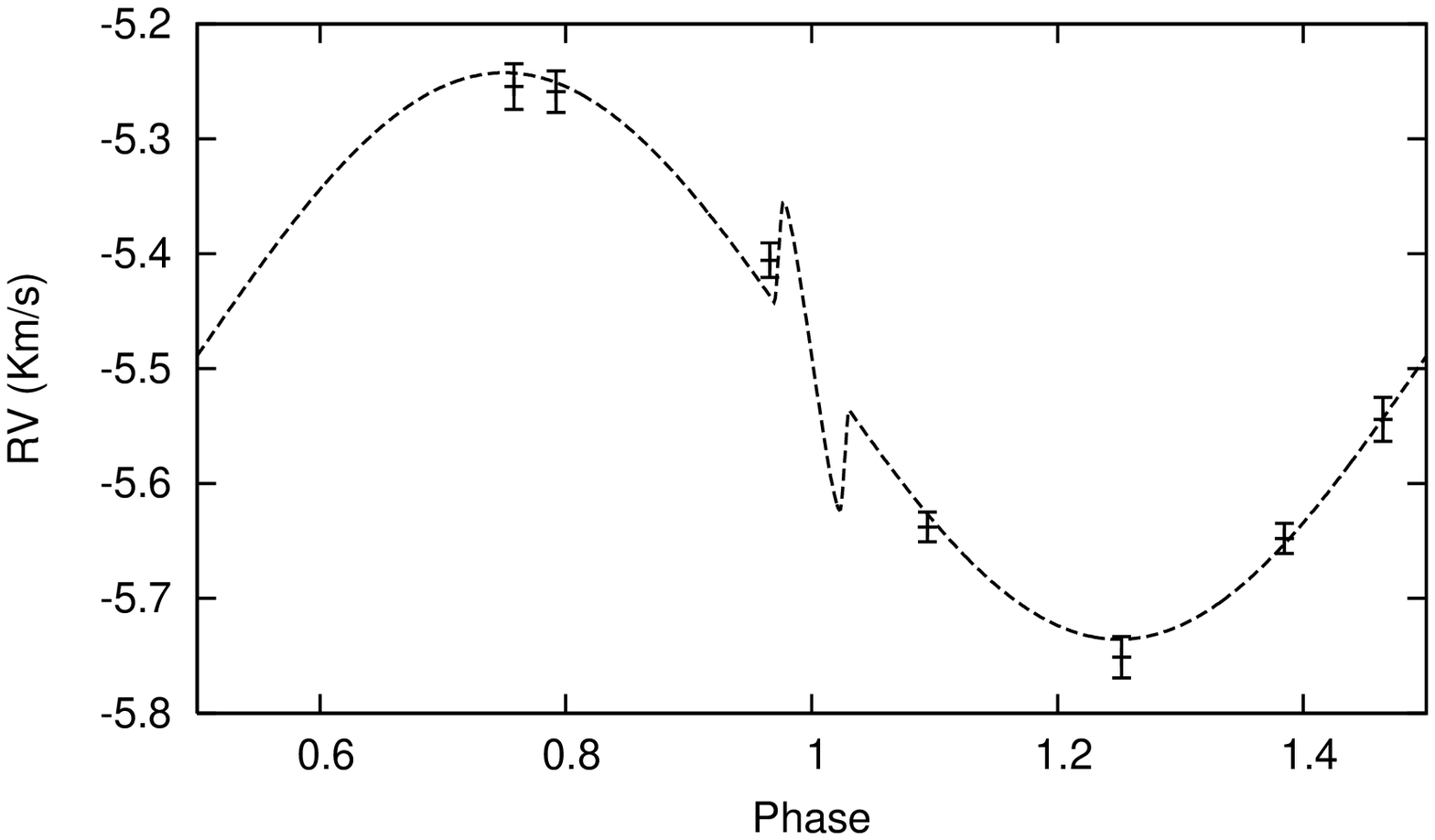,width=9.0cm,angle=0}
\caption[]{
Simultaneous MCMC solution to combined SuperWASP-N, IAC80 $V$, $I$ and Keele $R$ photometry. 
The orbital solution is assumed to be circular. The lower panel is the MCMC solution to the radial velocity data. The model here also shows the Rossiter-McLaughlin effect which is predicted to be significant in this system given the hosts $v sin i = 13.40$\,km/s.
}
\label{fig:sol_rvlc}
\end{center}
\end{figure}

The SOPHIE radial-velocity data measurements are plotted in Figure~\ref{fig:sol_rvlc} together with the best-fitting global fit  to the SuperWASP-N, IAC80 and Keele transit photometry. Since the timing of the transits and the radial-velocity solution both provide information about the orbit, we modelled the transit photometry and the reflex motion of the host star simultaneously.

The model of the primary star's radial-velocity orbit is parametrised in the usual way by the primary's radial-velocity amplitude $K_1$, the centre-of-mass velocity $\gamma$, the orbital eccentricity $e$ and the longitude $\omega$ of periastron.

The transit profile was modelled using the small-planet approximation of \citet{mandel2002}, with the 4-coefficient nonlinear limb-darkening model of \citet{claret2000}. We used $R$-band limb-darkening coefficients for the SuperWASP-N data, whose unfiltered wavelength response is centred near the $R$ band, and for the Keele $R$-band data. We used the appropriate $V$ and $I$-band limb-darkening coefficients for the IAC80 photometry. The transit model was characterised by the epoch $T_0$ of mid-transit, the orbital period $P$, the duration $t_T$ from first to fourth contact, the squared ratio $\Delta F = (R_p/R_*)^2$ of the planet radius $R_p$ to the stellar radius $R_*$, and the impact parameter $b=a(1-e\cos E_T)\cos i/R_*$ of the planet's trajectory across the face of the host star. Here $a$ is the orbital semi-major axis, $E_T$ is the eccentric anomaly at the epoch of transit and $i$ is the orbital inclination. The ratio of the stellar radius to the orbital separation is then given approximately (or exactly for a circular orbit) by
$$
\frac{R_*}{a}=\frac{t_T}{P}\frac{\pi}{(1+\sqrt{\Delta F})^2-b^2}
$$
\citep{cameron2007mcmc}. The orbital semi-major axis is derived from the orbital period and the stellar mass $M_*$ via Kepler's third law. The stellar mass is estimated from the $J-H$ colour as described by \citet{cameron2007mcmc}.

\begin{table*}
\caption[]{WASP-3 system parameters and 1-$\sigma$ error limits derived
from MCMC analysis.}
\label{tab:params}
\begin{tabular}{lccl}
\hline\\
Parameter & Symbol & Value & Units \\
\hline\\
Transit epoch (BJD) & $ T_0  $ & $ 2454143.8503 ^{+ 0.0004 }_{- 0.0003 } $ & days \\
Orbital period & $ P  $ & $ 1.846834 ^{+ 0.000002 }_{- 0.000002 } $ & days \\
Planet/star area ratio  & $ (R_p/R_s)^2 $ & $ 0.0106 ^{+ 0.0002 }_{- 0.0004 } $ &  \\
Transit duration & $ t_T $ & $ 0.1110 ^{+ 0.0009 }_{- 0.0018 } $ & days \\
Impact parameter & $ b $ & $ 0.505 ^{+ 0.051 }_{- 0.166 } $ & $R_*$ \\
  &    &      &  \\
Stellar reflex velocity & $ K_1 $ & $ 0.2512 ^{+ 0.0079 }_{- 0.0108 } $ & km s$^{-1}$ \\
Centre-of-mass velocity  & $ \gamma $ & $ -5.4887 ^{+ 0.0013 }_{- 0.0018 } $ & km s$^{-1}$ \\
Orbital semimajor axis & $ a $ & $ 0.0317 ^{+ 0.0005 }_{- 0.0010 } $ & AU \\
Orbital inclination & $ I $ & $ 84.4 ^{+ 2.1 }_{- 0.8 } $ & degrees \\
  &    &      &  \\
Stellar mass & $ M_* $ & $ 1.24 ^{+ 0.06 }_{- 0.11 } $ & $M_\odot$ \\
Stellar radius & $ R_* $ & $ 1.31 ^{+ 0.05 }_{- 0.12 } $ & $R_\odot$ \\
Stellar surface gravity & $ \log g_* $ & $ 4.30 ^{+ 0.07 }_{- 0.03 } $ & [cgs] \\
Stellar density & $ \rho_* $ & $ 0.55 ^{+ 0.15 }_{- 0.05 } $ & $\rho_\odot$ \\
  &    &      &  \\
Planet radius & $ R_p $ & $ 1.31 ^{+ 0.07 }_{- 0.14 } $ & $R_J$ \\
Planet mass & $ M_p $ & $ 1.76 ^{+ 0.08 }_{- 0.14 } $ & $M_J$ \\
Planetary surface gravity & $ \log g_p $ & $ 3.37 ^{+ 0.09 }_{- 0.04 } $ & [cgs] \\
Planet density & $ \rho_p $ & $ 0.78 ^{+ 0.28 }_{- 0.09 } $ & $\rho_J$ \\
Planet temp ($A=0$)  & $ T_{\mbox{eql}} $ & $ 1960 ^{+ 33 }_{- 76 } $ & K \\
\hline\\
\end{tabular}
\end{table*}

The set of nine parameters $\{T_0,P,t_T,\Delta F,b,M_*,K_1,e,\omega\}$ thus defines both the transit light curve and the form of the reflex velocity variation. We compute the photometric zero-point offset $\Delta m$ of the observed magnitudes $m_j$ from the model $\mu_j$ derived from a given set of parameters:
$$
\Delta m =\frac{ \sum_j (m_j-\mu_j)w_j}{\sum_j w_j}.
$$
The weights $w_j$ are the inverse variances $1/\sigma^2_j$ of the individual observations. Similarly, we compute the radial velocity $\gamma$ of the system's centre of mass as the inverse-variance weighted mean offset between the observed radial velocities $v_k$ and the model values $\nu_k$ for the current model parameters:
$$
\gamma =\frac{ \sum_k (v_k-\nu_k)w_k}{\sum_k w_k}.
$$

We quantify the goodness of fit to the data by the combined $\chi^2$ statistic for the combined photometric and radial-velocity data:
$$
\chi^2=\sum_{j=1}^{N_p}\frac{(m_j-\mu_j-\Delta m)^2}{\sigma^2_j}
+\sum_{k=1}^{N_v}\frac{(v_k-\nu_k-\gamma)^2}{\sigma^2_k}.
$$


Markov-Chain Monte-Carlo analysis has recently become established as an efficient and reliable method for establishing both photometric (\citealt{holman2006}; \citealt{burke2007xo2}) and orbital (\citealt{ford2006};  \citealt{gregory2007}) parameters of close-orbiting giant exoplanets. We determined the photometric and orbital parameters of the WASP-3 system using the Markov-chain Monte-Carlo algorithm described in detail by \citet{cameron2007mcmc}, to which we refer the reader for most details of the implementation. The initial photometric solution for the SuperWASP-N transit profiles is established by our transit-search algorithm \citep{c3}. The initial radial-velocity solution is an inverse variance-weighted linear least-squares fit assuming a circular orbit. In both cases, the initial fits also yield good estimates of the parameter uncertainties. The stellar mass is initialised at the value $M_0$ estimated from the $J-H$ colour.

At each step in the algorithm, each of the nine proposal parameters is perturbed by a small random amount such that
$$
T_{0,i} =  T_{0,i-1}+\sigma_{T_0}G(0,1)f
$$
and similarly for the other eight parameters. Here $G(0,1)$ is a random Gaussian deviate with mean zero and unit standard deviation. The adaptive step-size controller $f$ is initially set to 0.5, and evolves as the calculation progresses, ensuring that roughly 25 percent of proposal sets are accepted.

The prior probability distributions for all nine parameters are treated as being uniform. The parameters $P$, $t_T$, $\Delta F$, $M_*$ and $K_1$ are required to be positive. The impact parameter and eccentricity are 
restricted to the ranges $0<b<1$ and $0<e<1$, while the longitude of periastron is restricted to the range $-\pi<\omega<\pi$. The decision on whether or not to accept a set of proposal parameters is made via the Metropolis-Hastings algorithm using the logarithmic likelihood functional
$$
Q_i =\chi^2_i+\frac{(M_{*,i}- M_{0})^2}{\sigma^2_M}+\frac{(\log g_{*,i}- \log g_*)^2}{\sigma^2_{\log g}},
$$
where $\log g_{*,i}$ is computed directly from the mass $M_{*,i}$ and radius $R_{*,i}$. This imposes a Gaussian prior on the stellar mass,  and indirectly on the radius.  The prior forces the stellar mass to be close to the initial estimate $M_0$ with an assumed uncertainty $\sigma_M=0.1M_{0}$. The prior on 
$\log g_*$ ensures consistency with the spectroscopically-measured value $\log g_*=4.25\pm 0.05$, and thus helps to reduce the uncertainty in the stellar radius if the impact parameter is not strongly constrained by the photometry.

If a new set of proposal parameters yields $Q_i<Q_{i-1}$, the fit to the data is improved and the proposal is accepted. If $Q_i>Q_{i-1}$, the parameter set is accepted with probability $\exp[(Q_{i-1}-Q_i)/2]$. We find that the solution converges within a few hundred steps to a stable, optimal solution. After this initial "burn-in" phase, we re-scale the photometric error bars so that the contribution of each photometric dataset to $\chi^2$ is equal to the associated number of degrees of freedom. For the radial velocity data, we estimate the additional variance needed to match their contribution to $\chi^2$ with the number of degrees of freedom. This is equivalent to adding radial-velocity "jitter" with amplitude 10 m\,s$^{-1}$ in quadrature with the photon noise uncertainties giving the values listed in Table~\ref{tab:logspec}. We then run the algorithm for a few hundred  further steps and derive revised parameter uncertainties from the standard deviations of their respective Markov chains. This optimises the step length used in generating new sets of proposal parameters. Finally the algorithm is allowed to run for $10^5$ steps in order to map out the joint posterior probability distribution of the nine proposal parameters.

We find that for WASP-3 the nine proposal parameters  show only weak mutual correlations. The correlation lengths \citep{tegmark2004} of the Markov chains for individual parameters are typically 10 to 20 steps, so the final production run yielded approximately $10^4$ statistically independent parameter sets. In the initial runs we allowed all nine parameters to float, and arrived at a solution with eccentricity $e=0.05\pm 0.05$. Since this is statistically indistinguishable from the circular orbit expected for a planet with such a short period, the remaining eight parameters were fitted assuming $e=0$.

The values of the parameters at the optimal solution are given, together with their associated 1-$\sigma$ (68.3 percent) confidence intervals, in Table~\ref{tab:params}. These results are consistent with those derived from the spectral analysis presented earlier.

We explored the relationship between the stellar surface gravity and the impact parameter of the transit by repeating the fit for a sequence of values of $\log g$, with an artificially restricted $\sigma_{\log g}=0.01$. The results are given in Table~\ref{tab:logg}. The best formal fit to the photometry is obtained for stellar surface gravities in the range $4.35 < \log g < 4.45$. In this range, however, the stellar radius is unphysically low in relation to the stellar mass. At  the spectroscopically-determined $\log g = 4.25$, the photometric $\chi^2$ is only marginally degraded. At still lower values of $\log g$, the impact parameter 
increases to the point where the duration of transit ingress and egress cannot be fitted satisfactorily. Moreover, the radii of the star and the planet become implausibly inflated. We conclude that the stellar surface gravity must lie in the range $4.25 < \log g < 4.35$, and the impact parameter in the range 
$0.4 < b < 0.6$. The limits on these parameters derived from the full posterior probability distribution, as listed in Table~\ref{tab:params}, are consistent with this conclusion.

\begin{table}
\caption[]{Dependence of stellar and planetary parameters on $\log g_*$.}
\label{tab:logg}
\begin{tabular}{ccccccc}
\hline \\
$\log g$ & $\chi^2_{\mbox{ph}}$ & $b$ & $M_*$ & $R_*$ & $R_p$ & $\rho_p$ \\
(cgs)      &                        &  &($M_\odot$)&($R_\odot$)&($R_J$)&($\rho_J$) \\
\hline \\
4.05 & 4492.6 & 0.76 & 1.14 & 1.64 & 1.75 & 0.31 \\
4.15 & 4478.6 & 0.68 & 1.18 & 1.50 & 1.55 & 0.46 \\
4.25 & 4470.9 & 0.58 & 1.23 & 1.37 & 1.39 & 0.65 \\
4.35 & 4468.1 & 0.38 & 1.22 & 1.22 & 1.20 & 1.00 \\
4.45 & 4468.9 & $< 0.09$ & 1.36 & 1.17 & 1.14 & 1.25  \\
4.55 & 4491.1 & $< 0.05$ & 1.60 & 1.20 & 1.16 & 1.32 \\
\hline \\
\end{tabular}
\end{table}

\subsubsection{Line-bisector analysis}

\begin{figure}
\begin{center}
\psfig{figure=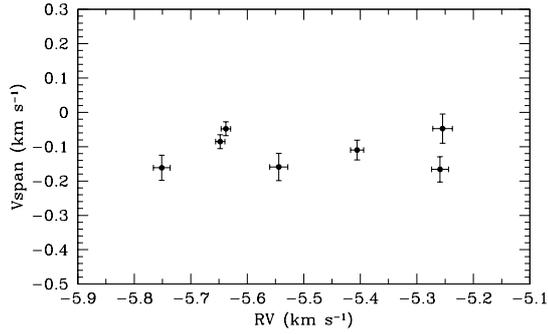,width=7.5cm}
\caption[]{Analysis of line-bisectors in WASP-3 shows the bisector velocity (vspan) does not correlate with stellar radial-velocity. This demonstrates that the cross-correlation function remains symmetric, and that the radial-velocity variations are not likely to be caused by line-of-site binarity or stellar activity.}
\label{fig:line_bisector}
\end{center}
\end{figure}

It is well known that faint binaries contaminating the photometric aperture, or even stellar  activity, can influence absorption line shape and can, in certain circumstances, mimic or confuse small radial velocity motions. By noting the position of the line-bisector of the cross-correlation function, 
asymmetries in the profile will become apparent. 

We measured the asymmetries of the cross-correlation function peaks using the line-bisector method as a function of radial velocity \citep{g1}, as applied by \citet{q1}. Figure~\ref{fig:line_bisector} demonstrates  that periodic variations, indicative of line-of-sight binarity or activity are not apparent, and we  conclude that the radial-velocity variations are genuinely due to to the orbital motion of a low mass object. 

\subsection{Adaptive Optics Imaging}

We further investigated the scenario of a triple system comprising  a bright single star and a faint, blended eclipsing-binary system  by performing high-resolution H-band imaging with the near-infrared camera INGRID, fed by the adaptive-optics system NAOMI, on the 4.2-m  William Herschel telescope. 
An image taken in natural seeing of 0.8\,arcsecond with corrected FWHM of 0.2\,arcsecond shows no evidence for resolved faint companions to WASP-3. Assuming an  F7-8V spectral type ($M_v \sim 3.8$) for WASP-3 would imply a distance of $\sim 220$ pc, hence these observations  constrain any potential 
associated eclipsing binary companion to lie within $\sim 45$ AU of the host.

\section{Discussion}

\begin{figure}
\begin{center}
\psfig{figure=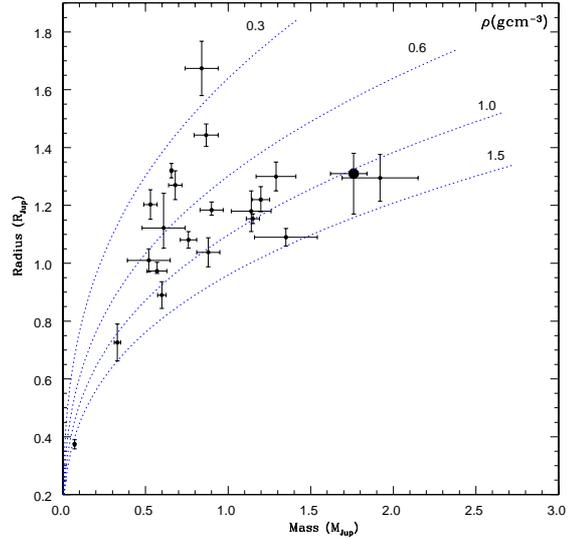,width=7.50cm,angle=0}
\caption[]{
The known confirmed transiting exoplanets plotted in the Mass-Radius plane.  Iso-density contours are plotted in cgs units. For clarity of scale we have not plotted  HD17156 (mass 3.12\,$M_J$, radius 1.15\,$R_J$) or the extremely high density object HD147506b (mass 8.04\,$M_J$, radius 0.98\,$R_J$).  WASP-3b is marked as a filled circle (data from  http://obswww.unige.ch/\~{}pont/TRANSITS.htm and references therein).}
\label{fig:mr}
\end{center}
\end{figure}

In this study we have found WASP-3b to be a transiting gas-giant exoplanet  with mass 
$ 1.76 ^{+ 0.08 }_{- 0.14 } M_J$  and radius $ 1.31 ^{+ 0.07 }_{- 0.14 } R_J$.  Its host star, WASP-3, has a photospheric temperature of $6400 \pm 100$\,K and $\log g = 4.25 \pm 0.05$, consistent with its F7-8V spectral type derived from 2MASS photometry. This places WASP-3b amongst the most massive of known transiting exoplanets (Figure~\ref{fig:mr}). Given the hosts relatively large rotational velocity and the large radius of the planet we would expect a significant amplitude for Rossiter-McLaughlin  effect (the model is included in Figure~\ref{fig:sol_rvlc}).

\begin{figure}
\begin{center}
\psfig{figure=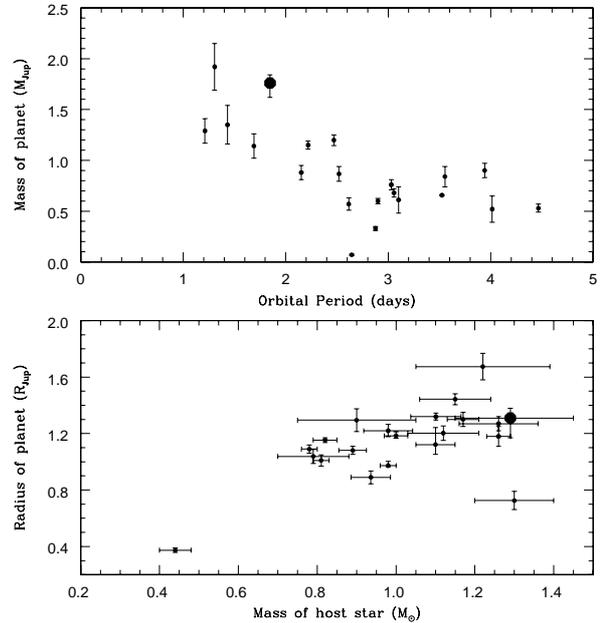,width=8.5cm,angle=0}
\caption[]{
Modified version of Figure 5 from \citet{sozz07}, incorporating a further seven newly discovered systems (data from  http://obswww.unige.ch/\~{}pont/TRANSITS.htm and references therein, but again excluding HD17156b and HD147506b). In each case WASP-3b is marked by the filled circle. 
}
\label{fig:planetparas}
\end{center}
\end{figure}

\begin{figure}
\begin{center}
\psfig{figure=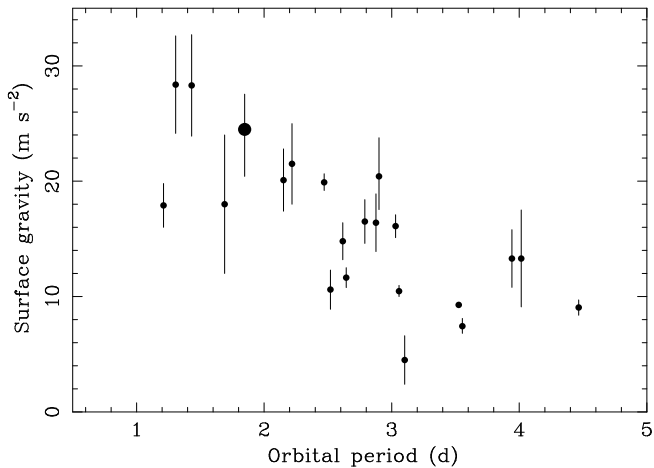,width=8.5cm,angle=0}
\caption[]{
The correlation of planetary surface gravity with orbital period for the 22 shortest period transiting planets(updated version from Figure~2 from \citet{southworth2007}).
}
\label{fig:planetgravity}
\end{center}
\end{figure}

\citet{sozz07} demonstrate a correlation of planet radius with host  mass for 14 confirmed transiting exoplanets, and a correlation of planet mass  with orbital period for the same sample (first noted by \citet{mazeh2005}). Figure~\ref{fig:planetparas} shows these relationships updated while in
Figure~\ref{fig:planetgravity} we also update the apparent correlation of surface 
gravity with orbital period noted by \citet{southworth2007}.  Despite the additional objects the 
$R_P$ v's $M_*$ correlation remains week (even ignoring the two most massive objects HD17156b and HD147506b). For both the $M_P$ v's $P$ and $g$ v's $P$ we contend that these relationships arise partly through observational selection and partly through the effects of the intense radiation fields that these planets are experiencing. We believe the absence of high gravity/mass bodies at  longer periods is primarily a detection effect, while the absence of low gravity/mass planets at short period could indeed be caused by rapid evaporation.

The closeness of the orbit and the large radius and high effective temperature of the star 
combine to make WASP-3b one of the most strongly-irradiated, and hence one of the hottest of the 
known exoplanets, second only to  OGLE-TR-56b and comparable to OGLE-TR-132. This raises 
the possibility that the atmosphere may be hot enough for TiO and VO to remain in the 
gas phase above the temperature minimum, creating a hot, strongly-absorbing stratosphere 
\citep{fortney2006hd149026b}. This would give an anomalously high infrared brightness 
temperature, as \citet{harrington2007hd149026b} inferred from {\em SPITZER/IRAC} 
secondary-eclipse photometry of  HD 149026b at 8$\mu$m. Being much closer and brighter 
than any of the OGLE host stars, WASP-3b is thus an excellent candidate for future 
observational tests of the hot-stratosphere hypothesis.

\section*{Acknowledgments}
The WASP Consortium consists of astronomers primarily from the Universities of Cambridge (Wide Field Astronomy Unit), Keele, Leicester, The Open University, Queen's University Belfast and St Andrews, the Isaac Newton Group (La Palma), the Instituto de  Astrof{\'i}sica de Canarias (Tenerife) and the South African Astronomical  Observatory. The SuperWASP-N and WASP-S Cameras were constructed and operated with funds made available from Consortium Universities  and the UK's Science and Technology Facilities Council (formerly PPARC). We extend our thanks to the Director and staff of the Isaac Newton Group of Telescopes and the South African Astronomical Observatory for their support of SuperWASP-N and WASP-S operations, and the Director and staff of the Observatoire de Haute-Provence for their support of the SOPHIE spectrograph.

\bsp

\label{lastpage}

\end{document}